\documentclass[twocolumn,superscriptaddress]{revtex4}
\usepackage{graphicx}

\begin{document}
\title{Separable metamaterials: analytical ab-initio homogenization and chirality}

\author{Alessandro Ciattoni}
\affiliation{Consiglio Nazionale delle Ricerche, CNR-SPIN, Via Vetoio 10, 67100 L'Aquila, Italy}

\author{Domenico Rago}
\affiliation{Dipartimento Scienze Fisiche e Chimiche, Universt\`a degli studi dell'Aquila, Via Vetoio, 67100 Coppito, L'Aquila, Italy}

\author{Carlo Rizza}
\affiliation{Dipartimento di Scienza e Alta Tecnologia, Universit\`a degli studi dell'Insubria, Via Valleggio 11, 22100 Como, Italy}
\affiliation{Consiglio Nazionale delle Ricerche, CNR-SPIN, Via Vetoio 10, 67100 L'Aquila, Italy}

\begin{abstract}
We investigate the ab-initio homogenization of separable metamaterials with factorized dielectric permittivity profile which can be achieved through suitable grey-scale permittivity design techniques. Separability allows such metamaterials to be physically regarded as the superposition of three fictitious 1D generating media. We prove that, in the long-wavelength limit, separable metamaterials admit simple and analytical description of their electromagnetic bi-anisotropic response which can be reconstructed from the properties of the 1D generating media. Our approach provides a strategy which allows the full ab-initio and flexible design of a complex bianisotropic response by using 1D metamaterials as basic building blocks.
\end{abstract}

\maketitle

Electromagnetic propagation through metamaterials is usually described by means of effective medium theories \cite{Smit1,Smit4,Felb1,Orti1} which are adequate when the radiation wavelength is much larger than the structure periodicity. Basically such homogenization approaches provide a method for evaluating the effective permittivity and permeability of the homogeneous effective medium from the knowledge of the electromagnetic properties of the metamaterial inclusions and their spatial arrangement. More refined approaches \cite{Silv1,Aluu1,Reye1,Ciat1}, encompassing the bianisotropic response of the inclusions and/or the spatial dispersion of the overall structure, allow to evaluate the bianisotropic effective parameters of the effective medium.

Chiral metamaterials, whose structure cannot be superimposed onto their mirror images, have attracted much research effort in the last decade \cite{Vale1,Liii1} since they host remarkable electromagnetic effects which are much more pronounced then in natural chiral media or cannot be observed in nature at all. Relevant examples of such phenomena are negative refraction \cite{Pend1,Plum1,Wang1,Zhan1,Liii2}, giant optical activity \cite{Plum2,Yeee1,Liuu1}, asymmetric transmission \cite{Menz1,Liuu2} and switching of the chiral metamaterial response \cite{Kann1,Kang1,Kena1}. Since chiral asymmetry is a 3D geometrical property, chiral metamaterials are generally composed of 3D metallic inclusions, e.g. twisted crosses or twisted split-ring resonators. Remarkably, an effective arbitrary reciprocal bi-anisotropic response can be tailored by suitably inserting in the metamaterial unit cell the basic chiral and omega inclusions \cite{Serdy}. Metamaterials exhibiting 2D and 1D geometric chirality, i.e. that cannot be superimposed onto their mirror images using only rigid motions of a plane (translations and in-plane rotation) and of a line (translations), have also been considered. Strong optical activity and circular dichroism have been predicted and observed in planar chiral metamaterials \cite{Papa1,Fedo1,Baii1,Deck1,Sing1,Gork1} and extrinsically chiral metamaterials \cite{Plum3,Plum4,Plum5,Caoo1,Luuu1,Huuu1,DeLe1,Shii1} which are characterized by 2D and 1D geometric chirality, respectively. Another class of metamaterials displaying 1D geometric chirality is that of 1D metamaterials \cite{Ciat1} which have recently been shown to exhibit highly tunability if containing graphene sheets \cite{Rizz1} and to support strong optical activity in the epsilon-near-zero regime \cite{Rizz2}.

In this paper we investigate a class of metamaterials characterized by a factorized and periodic microscopic permittivity profiles and we show that such \textit{separable metamaterials} can be analytically "decomposed" in terms of three fictitious 1D generating media. The decomposition is entailed by the factorization of the constituent permittivity profile as the product of three functions each depending on a single cartesian coordinate. The factorized permittivity profile can be obtained through a number of gray-scale permittivity design techniques, as all-photodesigning both in semiconductors \cite{Kamar} and in phase change materials \cite{Wangg} and digitalization \cite{Della}. We show that a remarkable consequence of the decomposition is the possibility to reconstruct the effective response (in the long-wavelength limit) from those of the 1D generating media. Therefore the description of the effective bianisotropic response, for separable metamaterials, is ab-initio, simple and fully analytical, and this allows the complete and simultaneous design of both dielectric and chiral properties.

\begin{figure*}
\center
\includegraphics[width=1\textwidth]{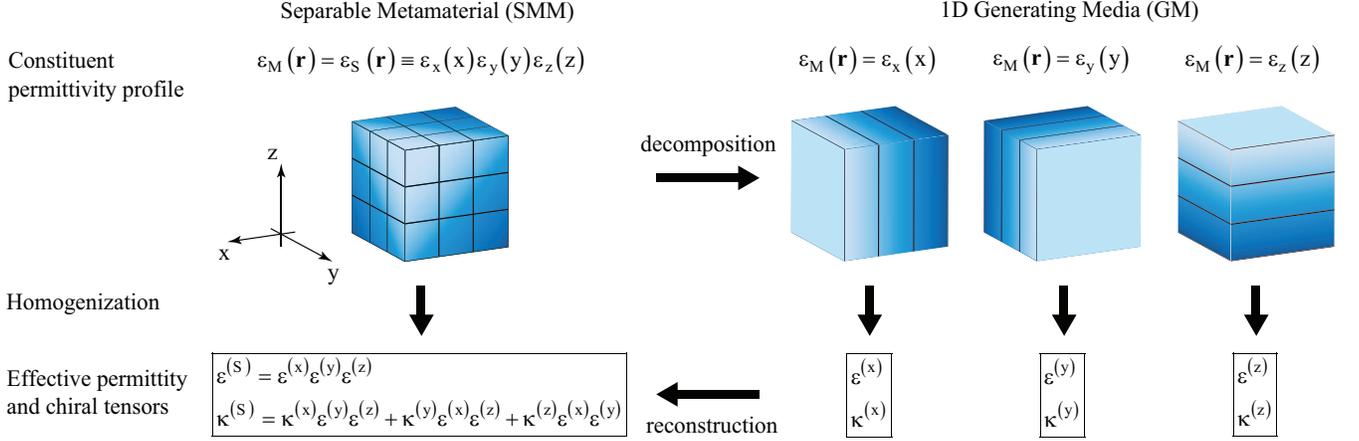}
\caption{(Color online) SMM homogenization can be broken into three steps: 1) decomposition into GM; 2) homogenization of the GM; 3) reconstruction of the SMM effective bianisotropy from those of the GM.}
\end{figure*}

Let us consider an unbounded metamaterial whose underlying nonmagnetic inclusions are spatially patterned, for simplicity, on a simple cubic lattice whose period $L$ is much smaller than the radiation wavelength $\lambda$, so that $\eta = L/\lambda \ll 1$. In this paper we consider the class of separable metamaterials (SMMs) whose microscopic relative dielectric permittivity $\varepsilon_S(\textbf{r})$ (the subscripts will hereafter be used for microscopic quantities) has the same periodicity of the metamaterial and it is separable, i.e. it can be written as
\begin{equation} \label{redu_eps}
\varepsilon_S (\textbf{r}) = \varepsilon_x (x) \varepsilon_y (y) \varepsilon_z (z),
\end{equation}
where $\varepsilon_\alpha(x_\alpha)$ ($\alpha=1,2,3$) are three one-dimensional periodic functions with period $L$. In order to obtain the effective electromagnetic response of SMMs in the long wavelength limit $\eta \ll 1$, we exploit the nonlocal homogenization approach theory developed in Ref.\cite{Ciat1}. According to such approach, a metamaterial with periodic microscopic dielectric constant $\epsilon_M(\textbf{r})$ (where the subscript $M$ labels an arbitrary medium) is described, up to the first order of $\eta$, by the effective medium constituent relations
\begin{eqnarray} \label{constitutive1}
\textbf{D} &=&  \varepsilon_0 \left(\varepsilon^{(M)} + \kappa^{(M)T} \kappa^{(M)} \right) \textbf{E} - \frac{i}{c} \kappa^{(M)T} \textbf{H},  \nonumber \\
\textbf{B} &=& \frac{i}{c} \kappa^{(M)} \textbf{E} + \mu_0 \textbf{H}.
\end{eqnarray}
Here $\varepsilon^{(M)}$ and $\kappa^{(M)}$ are the effective permittivity and chiral tensors (the superscripts will hereafter be used for effective quantities), respectively, and they are given by
\begin{eqnarray} \label{tensors}
\varepsilon_{\alpha \beta}^{(M)} &=&  \overline{\varepsilon_M} \delta_{\alpha \beta}+\frac{1}{2} \overline{\varepsilon_M \left(\partial_\alpha f_\beta + \partial_\beta f_\alpha \right) }, \nonumber \\
\kappa_{\alpha \beta}^{(M)} &=& k_0 \left[ -\epsilon_{\alpha \beta \mu} \overline{\varepsilon_M f_\mu} \right. \nonumber \\
&+& \left. \left(\epsilon_{\alpha \mu \nu} \delta_{\beta \rho} + \frac{1}{2} \delta_{\alpha \beta} \epsilon_{\mu \rho \nu} \right) \overline{\varepsilon_M f_\mu \partial_\rho f_\nu }\right],
\end{eqnarray}
where $k_0 = 2\pi /\lambda$, $\epsilon_{\alpha \beta \gamma}$ is the Levi-Civita symbol, the overline over a function $W$ stands for its spatial average over the cubic unit cell $C$, i.e. $\overline{W} = \frac{1}{L^3} \int_C d^3 \textbf{r} W (\textbf{r})$, and $f_\alpha (\textbf{r})$ are the three functions having the metamaterial periodicity with vanishing spatial average $(\overline{f_\alpha} = 0)$ which satisfy the equations
\begin{equation} \label{f-eq}
\nabla \cdot \left( \varepsilon_M \nabla f_\alpha \right) = - \partial_\alpha \varepsilon_M.
\end{equation}
Such electrostatic-like equation holds within the metamaterial unit-cell and the periodic boundary conditions on its edges, resulting from the metamaterial microscopic periodicity, physically accounts for the coupling among the unit cell and its surroundings. For a general dielectric profile $\varepsilon_M$, these equations have to be solved by resorting either to numerical or to spectral methods \cite{Felb1,Rizz3}. On the other hand, the situation of SMMs is peculiar since, for the dielectric profile of Eq.(\ref{redu_eps}) (i.e. for $\epsilon_M = \epsilon_R$), Eq.(\ref{f-eq}) yields the three independent equations
\begin{equation} \label{f-eq-1D}
\frac{d}{dx_\alpha} \left( \varepsilon_\alpha \frac{d f_\alpha}{dx_\alpha}  \right) = - \frac{d \varepsilon_\alpha}{dx_\alpha},
\end{equation}
in which the dependence $f_\alpha(\textbf{r}) = f_\alpha(x_\alpha)$ has been self-consistently assumed and which are satisfied by three functions $f_\alpha(x_\alpha)$ which are periodic with period $L$ and have vanishing spatial average.

It is now crucial noting that, for a periodic 1D medium whose dielectric permittivity is $\varepsilon_M(\textbf{r}) = \varepsilon_\alpha(x_\alpha)$, Eq.(\ref{f-eq}) reproduces Eq.(\ref{f-eq-1D}) \cite{Ciat1}. In other words the functions $f_x(x)$, $f_y(y)$ and $f_z(z)$ of a 3D SMM coincide with those pertaining three 1D media whose permittivities are $\varepsilon_x(x)$, $\varepsilon_y(y)$ and $\varepsilon_z(z)$. Therefore, it is natural to decompose a SMM into three fictitious 1D generating media (GM) whose permittivities are the $x_\alpha$-dependent parts of the separable permittivity profile of Eq.(\ref{redu_eps}). The effective permittivity and chiral tensors, $\varepsilon^{(\alpha)}$ and $\kappa^{(\alpha)}$, of the three GM are \cite{Ciat1}
\begin{eqnarray} \label{effec_1d}
\left(\varepsilon^{(\alpha)}\right)_{\beta \gamma} &=& \delta_{\beta \gamma} \left[\delta_{\beta \alpha} \frac{1}{b_{\alpha,0}} + \left( 1 - \delta_{\beta \alpha} \right) a_{\alpha,0} \right], \nonumber \\
\left(\kappa^{(\alpha)}\right)_{\beta \gamma} &=& \epsilon_{\beta \gamma \alpha} \frac{\tau_\alpha}{b_{\alpha,0}},
\end{eqnarray}
where $a_{\alpha,n} = \frac{1}{L} \int_0^L d \xi e^{-i 2\pi n (\xi/L) } \varepsilon_\alpha(\xi)$ and $b_{\alpha,n} = \frac{1}{L} \int_0^L d \xi e^{-i 2\pi n (\xi/L) } \left[\varepsilon_\alpha(\xi)\right]^{-1}$ are the Fourier coefficients of $\varepsilon_\alpha(x_\alpha)$ and its reciprocal, respectively, and $\tau_\alpha =-i \eta \sum_{n \neq 0} (a_{\alpha,-n} b_{\alpha,n})/n$. The parameter $\tau_\alpha$ is the degree of electromagnetic chirality of the GM associated to the $x_\alpha$ direction since it vanishes if the permittivity profile $\epsilon_\alpha (x_\alpha)$ can not be superposed onto its mirror image by using translations without resorting rotations (1D chirality) \cite{Ciat1,Rizz2}.

The above discussed decomposition of a SMM has remarkable consequences on its effective electromagnetic response. From Ref.\cite{Ciat1}, the functions $f_\alpha (x_\alpha)$ satisfying Eq.(\ref{f-eq-1D}) are $f_\alpha (x_\alpha) = \frac{L}{2\pi i b_{\alpha,0}} \sum_{n \neq 0} e^{-i 2\pi n (x_\alpha/L) } b_{\alpha,n}/n$ which inserted into Eqs.(\ref{tensors}), after some algebra, yield
\begin{eqnarray} \label{effec_tens}
\varepsilon^{(S)} &=& \left( \begin{array}{ccc}
                        \frac{a_{y,0} a_{z,0}}{b_{x,0}} & 0 & 0 \\
                        0 & \frac{a_{x,0} a_{z,0}}{b_{y,0}} & 0 \\
                        0 & 0 & \frac{a_{x,0} a_{y,0}}{b_{z,0}}
                       \end{array} \right), \nonumber \\
\kappa^{(S)}      &=& \left( \begin{array}{ccc}
                        0 & \frac{a_{x,0}}{b_{y,0}b_{z,0}} \tau_z & -\frac{a_{x,0}}{b_{y,0}b_{z,0}} \tau_y \\
                        -\frac{a_{y,0}}{b_{x,0}b_{z,0}} \tau_z & 0 & \frac{a_{y,0}}{b_{x,0}b_{z,0}} \tau_x \\
                        \frac{a_{z,0}}{b_{x,0}b_{y,0}} \tau_y & -\frac{a_{z,0}}{b_{x,0}b_{y,0}} \tau_x & 0
                       \end{array} \right),
\end{eqnarray}
expressions which are the main result of the present paper. From Eqs.(\ref{effec_tens}) it turns out that SMMs are generally biaxial media with pseudo-chiral-omega chirality ($\mathop{\rm Tr} \kappa^{(S)} =0$). Note that the homogenization procedure we have discussed to obtain these tensors from the constituent permittivity profile $\varepsilon_S ({\bf r})$ is particularly simple (i.e. it only requires the evaluation of the Fourier coefficients $a_{\alpha,n}$ and $b_{\alpha,n}$ and the summation of the series for $\tau_\alpha$). Moreover, Eqs.(\ref{effec_tens}) have a deeper conceptual value since, using Eqs.(\ref{effec_1d}), they can be rewritten as
\begin{eqnarray} \label{effec_tens2}
\varepsilon^{(S)} &=& \varepsilon^{(x)} \varepsilon^{(y)} \varepsilon^{(z)}, \nonumber \\
\kappa^{(S)}      &=& \kappa^{(x)} \varepsilon^{(y)} \varepsilon^{(z)} + \kappa^{(y)} \varepsilon^{(x)} \varepsilon^{(z)} + \kappa^{(z)} \varepsilon^{(x)} \varepsilon^{(y)}.
\end{eqnarray}
In other words SMMs have effective permittivity and chiral tensors which are combinations of the corresponding GM tensors, i.e. their bianisotropy is reconstructed from those of the GM. The whole analysis hitherto considered in summarized in Fig.1 where it is pictorially sketched that homogenization of a SMM can be broken into three steps: 1) decomposition into GM; 2) homogenization of the GM; 3) reconstruction of the SMM effective bianisotropy from those of the GM.

\begin{figure*}
\center
\includegraphics[width=1\textwidth]{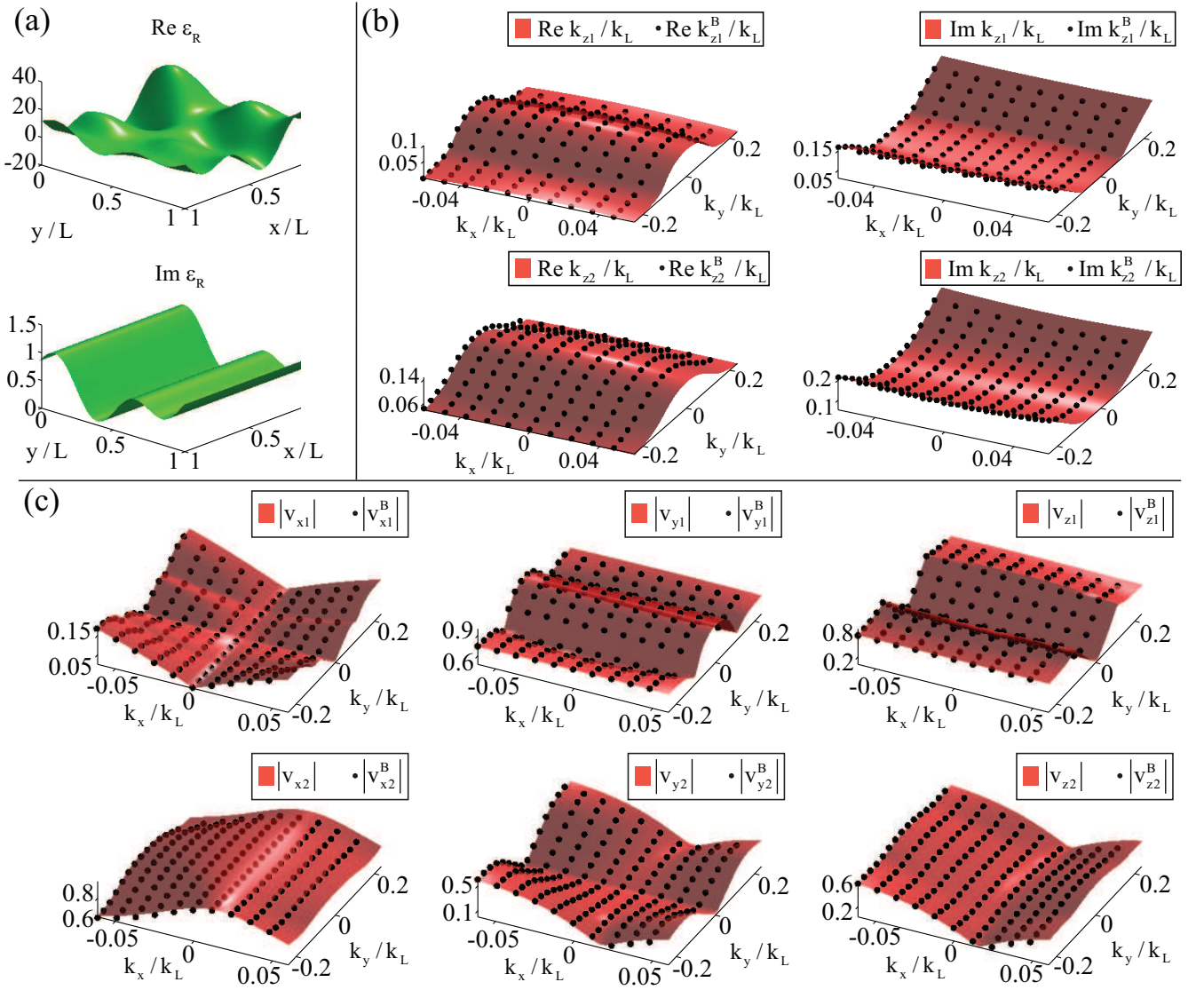}
\caption{(Color online) (a) Real and imaginary parts of the permittivity $\epsilon_R$ of the bi-dimensional SMM on the unit cell with $A_x = 1+0.2i$, $B_x = 2$, $A_y = 3$ and $B_y = 1.4$. (b) Real and imaginary parts of the effective medium eigenvalues $k_{z1}$, $k_{z2}$ (surfaces) and the Bloch eigenvalues $k_{z1}^B$, $k_{z2}^B$ (dots), normalized with $k_L = 2\pi /L$, for $\eta = L/\lambda = 0.07$. (c) Absolute values of the components of the normalized polarizations ${\bf v} = {\bf V} / |{\bf V} |$ and ${\bf v}^B = {\bf V}^B / |{\bf V}^B |$ of the modes of panel (b).}
\end{figure*}

In order to test the above results (Eq.(\ref{effec_tens2})) we have compared the isofrequency surface and polarizations of the bianisotropic effective medium with those predicted by the Bloch description of photonic crystals. For simplicity, we have focused on 2D SMMs whose constituent permittivity is $\varepsilon_S (x,y) = \epsilon_x(x) \epsilon_y(y)$ (i.e. with $\epsilon_z(z) = 1$) where
\begin{equation} \label{1D-example}
\epsilon_\alpha (x_\alpha) = A_\alpha + B_\alpha \left[ \cos \left( \frac{2\pi}{L} x_a \right) + \sin \left( \frac{4\pi}{L} x_a \right) \right].
\end{equation}
In Eq.(\ref{1D-example}), $A_\alpha = \overline{\epsilon_\alpha (x_\alpha)}$ whereas $B_\alpha$ is a parameter accounting for the 1D chirality of the profile $\epsilon_\alpha (x_\alpha)$ since, for $B_\alpha \neq 0$, there is no translation $T$ for which $\epsilon_\alpha (T - x_\alpha) = \epsilon_\alpha (x_\alpha)$. Maxwell equations for the effective medium of Eq.(\ref{constitutive1}), for the plane waves ${\bf E} = {\bf V} e^{i \left(k_x x + k_y y + k_z z\right)}$, turn into the homogeneous system $\left[ K^2 + i k_0 \left( K \kappa^{(S)} + \kappa^{(S)T} K \right) + k_0^2 \varepsilon^{(S)} \right]{\bf V}={\bf 0}$ where $(K)_{\alpha \beta} = \epsilon_{\alpha \beta \gamma} k_\gamma$, and $\varepsilon^{(S)}$ and $\kappa^{(S)}$ are the effective tensors of Eq.(\ref{effec_tens2}) evaluated from the permittivity profile $\varepsilon_S (x,y)$ through the above described procedure. Accordingly the eigenvalues $k_z \left(k_x ,k_y \right)$ and polarizations ${\bf V} \left(k_x ,k_y \right)$ are evaluated after imposing the compatibility of the system (i.e. after setting the determinant of the coefficient matrix equal to zero). On the other hand, in the Bloch description of photonic crystals, the periodic permittivity $\varepsilon_S (x,y)$ supports, for each $(k_x,k_y)$ in the first Brillouin zone $|k_x |< \pi/L$, $|k_y| <\pi/L$, the Bloch modes ${\bf E} = {\bf U}^{B}(x,y;k_z^B) e^{i \left(k_x x + k_y y + k_z^B z\right)}$ whose eigenvalues $k_z^B(k_x,k_y)$ and average polarizations ${\bf V}^B \left(k_x ,k_y \right) = \overline{{\bf U}^{B}(x,y;k_z^B)}$ are evaluated by solving Maxwell equations through the standard plane-wave-expansion method.

As a numerical example we consider the bi-dimensional SMM with $A_x = 1+0.2i$, $B_x = 2$, $A_y = 3$ and $B_y = 1.4$ and the period to wavelength ratio $\eta = 0.07$; in Fig.2(a) we plot the real and imaginary parts of its permittivity profile $\varepsilon_S$ on the unit cell. For each $(k_x,k_y)$, there are four eigenvalues $k_{zn}(k_x,k_y)$ in the effective medium description and an infinite number of eigenvalues $k_{zn}^B (k_x,k_y)$ in the Bloch description. In Fig.2(b) we plot the real and imaginary parts of the eigenvalues $k_{z1}$, $k_{z2}$ (surfaces) and $k_{z1}^B$, $k_{z2}^B$ (dots), normalized with $k_L = 2\pi /L$, where $k_{z1}$ and $k_{z2}$ are the effective medium eigenvalues with positive real parts whereas $k_{z1}^B$, $k_{z2}^B$ are the correspondingly closest Bloch eigenvalues. For each mode reported in Fig.2(b) we have evaluated the normalized polarizations ${\bf v} = {\bf V} / |{\bf V} |$ and ${\bf v}^B = {\bf V}^B / |{\bf V}^B |$ and we have plotted the absolute values of their components in Fig.2(c). The very good agreement between the two approaches proves the accuracy of the effective medium description of SMMs. Note, from the second row of Fig.2(c), that the modes $2$ have the polarization profiles which are not symmetric for the reflection $k_x \to -k_x$ and this is a clearly due to electromagnetic chirality which here plays a significant role.

An interesting feature of SMMs is that their bi-anisotropic response is produced by the full interplay of the three independent GM, since each of them affects all the components of the tensors of Eqs.(\ref{effec_tens}). Such remarkable property, once combined with the potential antiresonant $a_{\alpha,0} \simeq 0$ (Epsilon-Near-Zero regime \cite{Rizz2}) and resonant $b_{\alpha,0} \simeq 0$  behaviors of the GM, can be used to achieve enhancement and/or suppression of specific mechanisms. As an example we discuss the possibility of enhancing the medium electromagnetic chirality. Note that for a single 1D stratified medium, the condition $b_{\alpha,0} \simeq 0$ does not generally lead to the full enhancement of its electromagnetic chirality since $b_{\alpha,0}^{-1}$ appears in both tensors of Eqs.(\ref{effec_1d}). On the other hand, the interplay of the three GM can be used to compensate the resonance of one GM with the antiresonance of another GM to avoid large permittivities and, at the same time, to achieve large chirality tensors components. As an example, suppose that $b_{x,0} = \xi B_{x,0}$, $a_{y,0} = \xi A_{y,0}$ and $b_{z,0} = \xi B_{z,0}$, where $\xi \ll 1$ and the quantities labelled by capital letters are of the order of $1$. In this example the $x$- and $z$-associated GM are resonant whereas the $y$-associated GM medium is in the ENZ regime. In this situation the tensors of Eq.(\ref{effec_tens}) become
\begin{eqnarray} \label{effec_tens3}
\varepsilon^{(S)} &=& \left( \begin{array}{ccc}
                        \frac{A_{y,0} a_{z,0}}{B_{x,0}} & 0 & 0 \\
                        0 & \frac{a_{x,0} a_{z,0}}{b_{y,0}} & 0 \\
                        0 & 0 & \frac{a_{x,0} A_{y,0}}{B_{z,0}}
                       \end{array} \right), \nonumber \\
\kappa^{(S)}      &=& \frac{1}{\xi} \left( \begin{array}{ccc}
                        0 & \frac{a_{x,0}}{b_{y,0}B_{z,0}} \tau_z & -\frac{a_{x,0}}{b_{y,0}B_{z,0}} \tau_y \\
                        -\frac{A_{y,0}}{B_{x,0}B_{z,0}} \tau_z & 0 & \frac{A_{y,0}}{B_{x,0}B_{z,0}} \tau_x \\
                        \frac{a_{z,0}}{B_{x,0}b_{y,0}} \tau_y & -\frac{a_{z,0}}{B_{x,0}b_{y,0}} \tau_x & 0
                       \end{array} \right), \nonumber \\
\end{eqnarray}
so that, due to the factor $1/\xi \gg 1$, the components of the chirality tensor can become large while the permittivity do not experience enhancement.

The separability condition defining SMMs of Eq.(\ref{redu_eps}) is rather unusual and its achievement has to resort to specific expedients. A feasible way for achieving a factorized permittivity profile is offered by the recent photo-designing techniques where a general dielectric pattern is physically written within the medium bulk through suitable illumination. As an example, optically reconfigurable metasurfaces in phase change materials are achieved as a two-dimensional binary or greyscale pattern into a nanoscale thin film by inducing a refractive-index-changing phase transition with tailored trains of femtosecond pulses \cite{Wangg}. In order to show that a separable dielectric pattern can be achieved in phase change materials, we consider a slab of germanium antimony telluride ($Ge_3Sb_2Te_6$, or short GST) which has crystalline and amorphous phases whose dielectric permittivities, at $\lambda = 10 \: \mu m$, are $\epsilon_c = 38 + 2i$ and $\epsilon_a = 12.8 + 0.01i$, respectively \cite{Miche}. As shown in Ref.\cite{Wangg}, by using a spatial light modulator and femto-second pulses, it is possible to store a bidimensional periodic dielectric permittivity profile whose period is of the order of a micron. The structure of the unit cell can be suitably tailored since the local dielectric permittivity is given by $\epsilon_{loc} = \left[2 \epsilon_p - \epsilon_p^* + \sqrt{\left(2 \epsilon_p - \epsilon_p^*\right)^2 + 8 \epsilon_a \epsilon_c}\right] /4$ where $\epsilon_p = (1-f) \epsilon_a + f \epsilon_c$ and $\epsilon_p^* = (1-f) \epsilon_c + f \epsilon_a$ and $f$ is the local crystal fraction which can be related to the number of pulses locally striking the spot \cite{Wrigh}. This relation for $\epsilon_{loc}$ shows that $\textrm{Re} \left( \epsilon_{loc} \right)$ can be locally set to assume any values in the range between $\textrm{Re} \left( \epsilon_a \right)$ and $\textrm{Re} \left( \epsilon_b \right)$ and that its imaginary part can be neglected \cite{Miche}. As a basic and proof of concept example, one can design a basic unit cell comprising a bidimensional $2 \times 3$ rectangular array whose spots have light-induced permittivity $\epsilon_S ^{[ij]}$ where $i=(1,2)$ and $j = (1,2,3)$. The separability condition can be achieved by setting the values of the first and second row of the permittivity matrix as $\epsilon_S ^{[11]},\epsilon_S ^{[12]},\epsilon_S ^{[13]}$ and  $\theta \epsilon_S ^{[11]}, \theta \epsilon_S ^{[12]}, \theta \epsilon_S ^{[13]}$ (where $\theta$ is a parameter) in such a way that all the six values are in the above allowed light-induced permittivity range.

In conclusion, we have considered the novel class of SMMs whose permittivity separability yields a simple and interesting description of the medium electromagnetic response in the long-wavelength limit. Specifically we have shown that three one dimensional GM can be associated to a SMM and that the SMM bianisotropic response is obtained by combining those of the GM media. Both the permittivity and chiral tensor of a SMM have entries which are factorized as products of three contributions coming from the three 1D media. This provides the possibility of exploiting the interplay between the three GM to achieve, with desired suppressions and enhancements, the desired bianisotropic response. The most suitable platform for achieving and exploiting our results is naturally provided by schemes based on gray-scale permittivity design techniques, setups where the simple bianisotropy design offered by SMMs could suggest novel and pioneering devices for optical steering.

A Ciattoni and C Rizza thank the U.S. Army International Technology Center Atlantic for financial support (Grant No. W911NF-14-1-0315).


\begin{thebibliography}{99}

\bibitem{Smit1} D. R. Smith, S. Schultz, P. Markos and C. M. Soukoulis, Phys. Rev. B \textbf{65}, 195104 (2002).
\bibitem{Smit4} D. R. Smith, and J. B. Pendry, J. Opt. Soc. Am. B \textbf{23}, 391 (2006).
\bibitem{Felb1} D. Felbacq, G. Bouchitté, B. Guizal, and A. Moreau, J. Nanophoton. \textbf{2}, 023501 (2008).
\bibitem{Orti1} G. P. Ortiz, B. E. Martínez-Zérega, B. S. Mendoza, and W. L. Mochán, Phys. Rev. B \textbf{79}, 245132 (2009).
\bibitem{Silv1} M. G. Silveirinha, Phys. Rev. B \textbf{75}, 115104 (2007).
\bibitem{Aluu1} A. Alu, Phys. Rev. B \textbf{84}, 075153 (2011).
\bibitem{Reye1} J. A. Reyes-Avendaño, U. Algredo-Badillo, P. Halevi, and F. Pérez-Rodríguez, New J. Phys. \textbf{13}, 073041 (2011).
\bibitem{Ciat1} A. Ciattoni, and C. Rizza, Phys. Rev. B \textbf{91}, 184207 (2015).
\bibitem{Vale1} V. K. Valev, J. J. Baumberg, C. Sibilia, and T. Verbiest, Adv. Mater. \textbf{25}, 2517–2534 (2013).
\bibitem{Liii1} Z. Li, M. Mutlu, and E. Ozbay, J. Opt. \textbf{15}, 023001 (2013).
\bibitem{Pend1} J. B. Pendry, Science \textbf{306}, 1353 (2004).
\bibitem{Plum1} E. Plum, J. Zhou, J. Dong, V. A. Fedotov, T. Koschny, C. M. Soukoulis, and N. I. Zheludev, Phys. Rev. B \textbf{79}, 035407 (2009).
\bibitem{Wang1} B. Wang, J. Zhou, T. Koschny, and C. M. Soukoulis, App. Phys. Lett. \textbf{94}, 151112 (2009).
\bibitem{Zhan1} S. Zhang, Y. S. Park, J. Li, X. Lu, W. Zhang, and X. Zhang, Phys. Rev. Lett. \textbf{102}, 023901 (2009).
\bibitem{Liii2} Z. Li, R. Zhao, T. Koschny, M. Kafesaki, K. B. Alici, E. Colak, H. Caglayan, E. Ozbay, and C. M. Soukoulis, App. Phys. Lett. \textbf{97}, 081901 (2010).
\bibitem{Plum2} E. Plum, V. A. Fedotov, A. S. Schwanecke, N. I. Zheludev, and Y. Chen, App. Phys. Lett. \textbf{90}, 223113 (2007).
\bibitem{Yeee1} Y. Ye, and S. He, App. Phys. Lett. \textbf{96}, 203501 (2010).
\bibitem{Liuu1} M. Liu, D. A. Powell, I. V. Shadrivov, and Y. S. Kivshar, App. Phys. Lett. \textbf{100}, 111114 (2012).
\bibitem{Menz1} C. Menzel, C. Helgert, C. Rockstuhl, E.-B. Kley, A. Tunnermann, T. Pertsch, and F. Lederer, Phys. Rev. Lett. \textbf{104}, 253902 (2010).
\bibitem{Liuu2} D.-Y. Liu, X.-M. Zhai, L.-F. Yao, and J.-F. Dong, Opt. Commun. \textbf{323}, 19–22 (2014).
\bibitem{Kann1} T. Kan, A. Isozaki, N. Kanda, N. Nemoto, K. Konishi, H. Takahashi, M. Kuwata-Gonokami, K. Matsumoto, and I. Shimoyama, Nat. Commun. \textbf{6}, 8422 (2014).
\bibitem{Kang1} L. Kang, S. Lan, Y. Cui, S. P. Rodrigues, Y. Liu, D. H. Werner, and W. Cai, Adv. Mater. \textbf{27}, 4377–4383 (2015).
\bibitem{Kena1} G. Kenanakis, R. Zhao, N. Katsarakis, M. Kafesaki, C. M. Soukoulis, and E. N. Economou, Opt. Express \textbf{22}, 12149-12159 (2014).
\bibitem{Serdy} A. Serdyukov, I. Semchenko, S. Tretyakov, and A. Sihvola \textit{Electromagnetics of Bi-anisotropic Materials: Theory and Applications} (Gordon and Breach Science Publishers, Amsterdam, 2001).
\bibitem{Papa1} A. Papakostas, A. Potts, D.M. Bagnall, S. L. Prosvirnin, H. J. Coles, and N. I. Zheludev, Phys. Rev. Lett. \textbf{90}, 107404 (2003).
\bibitem{Fedo1} V. A. Fedotov, P. L. Mladyonov, S. L. Prosvirnin, A.V. Rogacheva, Y. Chen, and N. I. Zheludev, Phys. Rev. Lett. \textbf{97}, 167401 (2006).
\bibitem{Baii1} B. Bai, Y. Svirko, J. Turunen, and T. Vallius, Phys. Rev. A \textbf{76}, 023811 (2007).
\bibitem{Deck1} M. Decker, M. W. Klein, M. Wegener, and S. Linden, Opt. Lett. \textbf{32}, 856 (2007).
\bibitem{Sing1} R. Singh, E. Plum, C. Menzel, C. Rockstuhl, A. K. Azad, R. A. Cheville, F. Lederer, W. Zhang, and N. I. Zheludev, Phys. Rev. B \textbf{80}, 153104 (2009).
\bibitem{Gork1} M. V. Gorkunov, A. A. Ezhov, V. V. Artemov, O. Y. Rogov, and S. G. Yudin, App. Phys. Lett. \textbf{104}, 221102 (2014).
\bibitem{Plum3} E. Plum, V. A. Fedotov, and N. I. Zheludev, App. Phys. Lett. \textbf{93}, 191911 (2008).
\bibitem{Plum4} E. Plum, V. A. Fedotov, and N. I. Zheludev, J. Opt. A: Pure Appl. Opt. \textbf{11}, 074009 (2009).
\bibitem{Plum5} E. Plum, X.-X. Liu, V. A. Fedotov, Y. Chen, D. P. Tsai, and N. I. Zheludev, Phys. Rev. Lett. \textbf{102}, 113902 (2009).
\bibitem{Caoo1} T. Cao, C. Wei, L. Mao, and Y. Li, Sci. Rep. \textbf{4}, 7442 (2014).
\bibitem{Luuu1} X. Lu, J. Wu, Q. Zhu, J. Zhao, Q. Wang, L. Zhanb, and W. Ni, Nanoscale \textbf{6}, 14244–14253 (2014).
\bibitem{Huuu1} L. Hu, Y. Huang, L. Fang, G. Chen, H. Wei, and Y. Fang, Sci. Rep. \textbf{5}, 16069 (2015).
\bibitem{DeLe1} I. De Leon, M. J. Horton, S. A. Schulz, J. Upham, P. Banzer, and  R. W. Boyd, Sci. Rep. \textbf{5}, 13034 (2015).
\bibitem{Shii1} J. H. Shi, Q. C. Shi, Y. X. Li, G. Y. Nie, C. Y. Guan, and  T. J. Cui, Sci. Rep. \textbf{5}, 16666 (2015).
\bibitem{Rizz1} C. Rizza, E. Palange, and A, Ciattoni, Phot. Res. \textbf{2}, 121 (2014).
\bibitem{Rizz2} C. Rizza, A. Di Falco, M. Scalora, and A. Ciattoni, Phys. Rev. Lett. \textbf{115}, 057401 (2015).
\bibitem{Kamar} N. Kamaraju, A. Rubano, L. Jian, S. Saha, T. Venkatesan, J. N\"{o}tzold, R. K. Campen, M. Wolf and T. Kampfrath, Light: Science \& Applications {\bf 3}, e155 (2014).
\bibitem{Wangg} Q. Wang, E. T. F. Rogers, B. Gholipour, C.-M. Wang, G. Yuan, J. Teng, and N. I. Zheludev, Nature Photon. \text{10}, 60–65 (2016).
\bibitem{Della} C. Della Giovampaola and Nader Engheta, Nature Mater. \textbf{13}, 1115–1121 (2014).
\bibitem{Rizz3} C. Rizza, and A. Ciattoni, Photonics \textbf{2}, 365-374 (2015).
\bibitem{Miche} A. K. U. Michel, D. N. Chigrin, T. W. W. Maß, K. Schönauer, M. Salinga, M. Wuttig, and T. Taubner, Nano Lett. \textbf{13}, 3470–3475 (2013).
\bibitem{Wrigh} C. D. Wright, Y. Liu, K. I. Kohary, M. M. Aziz, and R. J. Hicken, Adv. Mater. \textbf{23}, 3408-3413 (2011).
\end{thebibliography}
\end{document}